\documentclass[
reprint,
nofootinbib,
amsmath,amssymb,amsbsy,
prd,
twocolumn,
superscriptaddress,
longbibliography,
preprintnumbers]{revtex4-2}

\usepackage{relsize}

\usepackage[acronyms, nohypertypes={acronym}, nopostdot, style=super, nonumberlist, toc]{glossaries}
\glsenableentrycount
\setacronymstyle{long-sc-short}

\newcommand\ac[1]{\gls{#1}}

\newacronym{WF}{wf}{Wilson-Fisher}
\newacronym{AF}{af}{asymptotically free}
\newacronym{RG}{rg}{renormalization group}
\newacronym{WZW}{wzw}{Wess-Zumino-Witten}
\newacronym[longplural={conformal field theories}]{CFT}{cft}{conformal field theory}
\newacronym[longplural={lattice field theories}]{LFT}{lft}{lattice field theory}
\newacronym[longplural={effective field theories}]{EFT}{eft}{effective field theory}
\newacronym[longplural={quantum field theories}]{QFT}{qft}{quantum field theory}
\newacronym{LEC}{lec}{low-energy constant}
\newacronym{QCD}{qcd}{quantum chromodynamics}
\newacronym{MC}{mc}{Monte Carlo}
\newacronym{IR}{ir}{infrared}
\newacronym{UV}{uv}{ultraviolet}
\newacronym{SNR}{snr}{signal-to-noise ratio}
\newacronym{NLSM}{nl$\sigma$m}{nonlinear sigma model}
\newacronym{PCM}{pcm}{principal chiral model}
\newacronym{CSA}{csa}{Cartan subalgebra}
\newacronym{SSB}{ssb}{spontaneous symmetry breaking}
\newacronym{DOF}{dof}{degrees of freedom}
\newacronym{DMRG}{dmrg}{densiy matrix renormalization group}
\newacronym{YM}{ym}{Yang-Mills}
\newacronym{QLM}{qlm}{quantum link model}
\newacronym{KG}{kg}{Kogut-Susskind}
\newacronym{KG-QLM}{kg-qlm}{Kogut-Susskind quantum link model}
\newacronym{D-QLM}{d-qlm}{D-theory quantum link model}
\newacronym{SPT}{spt}{symmetry protected topological} 
\newacronym{GW}{gw}{Ginsparg-Wilson}
\newacronym{FK}{fk}{Fidkowski-Kitaev}
\newacronym{CS}{cs}{Chern-Simons}
\newacronym{APS}{aps}{Atiyah-Patodi-Singer}
\newacronym{PV}{pv}{Pauli-Villars}

\usepackage{mathtools}
\usepackage{amsmath,amssymb,amsbsy,amsfonts}

\usepackage{hyperref}
\hypersetup{
    unicode,
    pdfprintscaling=None,
    colorlinks,
    breaklinks
}
\hypersetup{
  colorlinks=true,   
  linkcolor=cyan,    
  citecolor=blue,    
  filecolor=magenta, 
  urlcolor=blue      
}

\usepackage[capitalise]{cleveref}
\crefname{section}{Sec.}{Secs.}

\newcommand\Order{O}

\newcommand\sqvpM{\sqrt[+]{-\Vmaj}}

\newcommand\sqvmM{\sqrt[-]{-\Vmaj}}

\usepackage{microtype}

\newcommand\del\partial

\newcommand\Tsym{\ensuremath{\textsf{T}}}

\newcommand\ZZ{\mathbb{Z}}
\usepackage{dutchcal}

\newcommand\tsum{\textstyle\sum}

\newcommand\Dirac{\slashed{D}}
\newcommand\Dov{\CD_{\mathrm{ov}}}
\newcommand\Dgw{\CD_{\mathrm{GW}}}
\newcommand\Dpv{\CD_{\mathrm{PV}}}

\newcommand\mbf{ \mathsf{m}}
\newcommand\Dbf{ \mathsf{D}}

\newcommand\Dbfpv{\Dbf_{\mathrm{PV}}}
\newcommand\Dbfov{\Dbf_{\mathrm{ov}}}

\newcommand\Abf{\mathsf{A} }
\newcommand\Bbf{\mathsf{B} }

\newcommand\Anomaly{\mathcal{A}}

\newcommand\Cmat{\mathcal{C}}
\newcommand\Tmat{\mathcal{T}}

\usepackage{makecell}
\newcommand\TopRule{\Xhline{0.08em}}

\newcommand\MidRule{\Xhline{0.03em}}
\newcommand\BotRule{\Xhline{0.08em}}

\newcommand\Cm{{\mathcal{M}}}
\newcommand\Ck{\mathcal{K}}

\newcommand\Dw{D_\text{w}}
\newcommand\Dwm{\tilde{D}_\text{w}}

\newcommand\Qw{Q_\omega}
\newcommand\Xw{X_\omega}
\newcommand\Xwbf{{\mathsf X}_\omega}
\newcommand\Deltaw{\mbf_\omega^{-1} \mbf}

\newcommand\Rsym{\mathsf{R}}

\newcommand\Nf{N_{F}}

\newcommand\gammachi{\bar{\gamma}}

\usepackage{enumitem}
\usepackage{pifont}
\newcommand{\xmark}{\ding{55}}%

\usepackage{graphicx}
\usepackage{amsfonts}
\usepackage[named]{xcolor}
\usepackage{tocvsec2}
\usepackage{slashed}
\usepackage{cancel}
\usepackage{comment}
\newcommand{\chck}{\pmb{\checkmark}}

\newcommand\beq{\begin{eqnarray}}
\newcommand\eeq{\end{eqnarray}}

\newcommand\Dsl {\slashed{D}}
\newcommand{\half}{\frac{1}{2}}

\newcommand\eqn[1]{\label{eq:#1}}

\newcommand{\bfx}{{\mathbf x}}

\newcommand{\bfm}{{\mathbf m}}
\newcommand{\bfn}{{\mathbf n}}

\newcommand{\CA}{{\cal A}}

\newcommand{\CC}{{\cal C}}
\newcommand{\CD}{{\cal D}}

\newcommand{\CK}{{\cal K}}

\newcommand{\CM}{{\cal M}}

\newcommand{\CR}{{\cal R}}

\newcommand{\CT}{{\cal T}}
\newcommand{\CU}{{\cal U}}

\newcommand{\Tr}{{\rm Tr\,}}

\newcommand{\mybar}[1]{\kern 0.6pt\overline{\kern -0.6pt#1\kern -0.6pt}\kern 0.6pt}

\def\Tr{\text{Tr}\,}

\def\half{\tfrac{1}{2}}

\def\U\Omega{U(1)_{\Omega}}

\def\Vmaj{V_{\rm maj}}

\usepackage[section]{placeins}

\begin{document}

\preprint{INT-PUB-23-024,IQuS@UW-21-064,FERMILAB-PUB-23-541-T}
\title{Generalized Ginsparg-Wilson relations}

\author{Michael Clancy}
\email{mclancy2@uw.edu}
\affiliation{Institute for Nuclear Theory, Box 351550, Seattle WA 98195-1550}

\author{David B. Kaplan}
\email{dbkaplan@uw.edu}
\affiliation{Institute for Nuclear Theory, Box 351550, Seattle WA 98195-1550}

\author{Hersh Singh}
\email{hershs@fnal.gov}
\affiliation{Fermi National Accelerator Laboratory, Batavia, Illinois, 60510, USA}

\begin{abstract}
We give a general derivation of Ginsparg-Wilson relations for both Dirac and Majorana fermions in any dimension.  These relations encode continuous and discrete chiral, parity and time reversal anomalies and will apply to the various classes of free fermion topological insulators and superconductors (in the framework of a relativistic quantum field theory in Euclidean spacetime). We show how to formulate the exact symmetries of the lattice action and the relevant index theorems for the anomalies.
\end{abstract}

\maketitle

\section{Introduction}
\label{sec:intro}

The \ac{GW} relations govern how massless lattice fermions without doublers can optimally realize anomalous continuum symmetries \cite{Ginsparg:1981bj,Neuberger:1997fp,hasenfratz1998index,Luscher:1998pqa}.  They were originally derived for describing massless Dirac fermions with  chiral symmetries in even  spacetime dimensions, while analogous relations were posited for a massless Dirac fermion in three dimension with a parity anomaly \cite{bietenholz2001ginsparg}.  Lattice operators which satisfy these relations realize anomalous symmetries in the ``best'' possible way: the fermion propagator respects the symmetry at any nonzero spacetime separation, and as in the continuum, the lattice action possesses an exact, nearly local form of the symmetry  \cite{Luscher:1998pqa}, which is therefore respected by the Feynman rules in perturbative calculations.  On the other hand, the lattice integration measure is not invariant under this ``L\"uscher symmetry'', and the resultant Jacobian in the lattice theory correctly reproduces  the continuum anomaly expressed in terms of the index of the fermion operator. Here we give a unified derivation of such relations for Dirac and Majorana fermions alike in any dimension, and show how these continuous and discrete anomalous symmetries are realized. 
The connection between \ac{GW} fermions and extra dimensions is well established --- the first explicit solution to the \ac{GW} equations being the overlap operator \cite{Neuberger:1997fp,Narayanan:1993zzh,Narayanan:1993ss,Narayanan:1993sk,neuberger1998vectorlike} which was derived  to describe edge states of domain wall fermions in one higher dimension \cite{Kaplan:1992bt,Jansen:1992tw,Golterman:1992ub,kaplan1996domain}. It has since been understood that these relativistic systems are equivalent to the  topological insulators and superconductors studied in condensed matter physics, and so the generalized \ac{GW} relations we derive  apply to the massless edge states of the wide variety of topological classes \cite{schnyder2008classification,Kitaev:2009mg} of such materials\footnote{Soon after this paper appeared on the arXiv, another work appeared which discusses the topological classes of relativistic lattice fermions in detail, along with their \ac{GW} relations \cite{Kimura:2023kqs}.}.

 In the following analysis we are interested in the cases of $\Nf$ flavors of Dirac or Majorana fermions where
(i) the massless theory respects a symmetry $G$;
(ii) a mass term is possible for regulating the theory;
(iii) the mass term necessarily breaks the symmetry $G$.
In this set of circumstances we expect the massless theory to have an t'Hooft anomaly involving the $G$ symmetry, a \ac{GW} relation to exist for the ideally regulated fermion operator, and the existence of an exact $G$ symmetry obeyed by the regulated action, for which the Jacobian  reproduces the anomaly of the continuum theory, a generalization of  L\"uscher symmetry\footnote{Notation: we use upper case Greek letters such as $\Psi(x)$ to denote continuum fields, and lower case, such as $\psi_\bfn$ for lattice variables,  generally suppressing  indices for the latter. We take Euclidean $\gamma$ matrices to be Hermitian with $\{ \gamma^\mu, \gamma^\nu \} = 2 g^{\mu \nu}$; the gauge covariant Dirac operator $\Dirac = \gamma_\mu D_\mu$ is therefore anti-Hermitian with imaginary eigenvalues.  For a regulated Dirac operator, such as a generic Ginsparg-Wilson operator, overlap operator, or Pauli-Villars regulated operator, we use the notation $\Dgw, \Dov, \Dpv$ or simply $\cal{D}$.
  For Majorana fermions, we work with an antisymmetric kinetic and mass operators denoted as $\Dbf$ and $\mbf$.
  We use the mostly plus convention for our Minkowski metric. }. 

\section {Generalized Ginsparg-Wilson relations for Dirac fermions}
\label{sec:gw-dirac}

\subsection{Derivation of the relations}
\label{sec:gw-dirac-derivation}

Following the logic of the original derivation, we start by considering the  continuum theory of a free Dirac fermion $\Psi$  in Euclidean spacetime of arbitrary dimension, possibly in  background gauge or gravitational fields, described by the path integral
\begin{align}
  Z = \int d\Psi\, d\bar\Psi e^{-S(\bar\Psi,\Psi)}\ .
\end{align}
We now do a block transformation, defining a function $f(\bfx)$ whose support lies in a volume  $a^d$ about the origin, and our block averaged variables to be
\begin{align}
  \psi_\bfn = \int  d^d\bfx\ \Psi(\bfx) f(\bfx-\bfn a)\,
  \eqn{discrete}
\end{align}
and similarly for $\bar\psi_\bfn$. The parameter $a$ will be our lattice spacing, and for the rest of this article we will work in ``lattice units'' with $a=1$.
Lattice variables $\chi_\bfn$ and a lattice action $S_\text{lat} = \bar\chi \CD \chi$ are defined by
\begin{align}
  e^{-\bar\chi \CD \chi} =  \int d\Psi\, d\bar\Psi e^{-S(\bar\Psi,\Psi)}\,e^{-  (\bar\psi -\bar \chi )   m  ( \psi - \chi ) }
  \eqn{latact}\end{align}
so that up to an overall normalization, 
\begin{align}
  Z = \int \prod_n d\chi  d\bar\chi \, e^{-\bar\chi \CD \chi} \ .
\end{align}
The parameter $ m$ is an invertible Hermitian matrix which we can take to be a real number $ m$ times the identity matrix, but we will leave it in matrix form for now so that the identities for Dirac fermions and Majorana fermions (for which $ m$   is replaced by $\mbf$, an imaginary antisymmetric matrix) look similar.

We now assume that the continuum action $S$ is invariant under a global symmetry  transformation $\Psi \to \Omega \Psi$, $\bar\Psi\to \bar\Psi \bar\Omega $, where $\bar\Omega$ and $\Omega$ are some operators.  The symmetry transformations of interest are those which are broken by the Gaussian term  proportional to $ m$ that we have added to the path integral.   Examples we will consider include a $U(1)_A$ chiral transformation, a discrete chiral transformation (not contained in $U(1)_A$), and a coordinate reflection:
\begin{align}
  \Omega &= \bar \Omega = e^{i\alpha \gammachi}\quad &\text{(chiral symmetry)}\ ,\eqn{chiral}\\
  \Omega &= \bar \Omega =   \gammachi\quad &\text{(discrete chiral symmetry)}\ ,\eqn{discretechiral}\\
    \Omega  &= -\bar \Omega = \varepsilon\CR_1\gamma_1\quad &\text{(reflection symmetry)}\ ,\eqn{reflect}
\end{align}
with $\gammachi$ being the analog of $\gamma_5$ in arbitrary even dimension, 
where $\CR_1$ reflects the sign of the $x_1$ coordinate; generally $\varepsilon=1$, but in certain Majorana theories  $\varepsilon =i$.  Under reflections we assume that background fields are similarly reflected.   We will subsequently  consider an antilinear symmetry in Euclidean space related to time reversal in Minkowski spacetime.  We focus primarily on a single flavor of fermion, and hence do not discuss nonabelian flavor symmetries, but our analysis can be easily extended to include those. Other symmetries which are directly broken by the discretization function $f$, such as translation symmetry, spacetime rotations, conformal transformations or supersymmetry transformations do not seem to yield useful relations and we do not consider these (see \cite{bergner_blockinginspired_2013,bergner_generalising_2009} for interesting attempts in these directions).  

While the action  is invariant under the $\Omega,\bar\Omega$ transformation, the measure generally transforms as $d\Psi d\bar\Psi\to d\Psi d\bar\Psi \, e^{2i\CA}$, where $\CA$ is called the anomaly and arises from the Jacobian of the transformation \cite{fujikawa1979path}.   

We  wish to distinguish between the continuum transformation $\Omega$ and the transformation $\omega$ of the block averaged variables,
\begin{align}
  \psi_\bfm \to \int \Omega \,\Psi(\bfx) f(\bfx-a\bfm)\, d^d\bfx = \omega_{\bfm\bfn}\, \psi_\bfn\ .
  \eqn{omdef}
\end{align}
The matrices $\omega,\bar\omega$ are the lattice regulated forms of  $\Omega,\bar \Omega$.  They act as ordinary matrices on the lattice variables $\psi_\bfn$, but in the case of reflections, they also reflect the background fields.
Defining 
\begin{align}
  \CD_\omega = \bar \omega \CD \omega\ ,\quad  m_\omega=\bar\omega  m \omega\ ,
  \eqn{GW0}
\end{align}
it follows that
\begin{align}
  e^{-\bar\chi \CD_\omega \chi} =    \int d\Psi\, d\bar\Psi \,e^{2i\CA}\,e^{-S(\bar\Psi,\Psi)}\,e^{-(\bar\psi -\bar \chi )   m_\omega    ( \psi - \chi) }\ .
  \eqn{transformedPI}
\end{align}
Using the relation \cref{eq:Diracident} we have
\begin{align}
   e^{-  (\bar\psi  -\bar \chi  )  m_\omega  ( \psi  - \chi  ) } &= e^{\Tr\ln m_\omega m^{-1} }\,e^{  \partial_\chi  \Xw \partial_{\bar\chi}} \, e^{-   (\bar\psi  -\bar \chi  )   m ( \psi  - \chi  ) }  , 
\end{align}
where
\begin{align}
  \quad 
  \Xw  = m^{-1} - m_\omega^{-1}\ ,
\end{align}
and so 
\begin{align}
  e^{-\bar\chi \CD_\omega \chi}
   &= e^{2i\CA}\,e^{\Tr\ln m_\omega m^{-1}}\,e^{  \partial_\chi \Xw \partial_{\bar\chi}} \,e^{-\bar\chi \CD \chi} \nonumber\\
    &= e^{2i\CA}\,e^{\Tr\ln m_\omega m^{-1} +\Tr\ln \Qw}\, e^{-\bar\chi \frac{1}{\Qw}\CD \chi}\ ,
\label{eq:master}
\end{align}
where
\begin{align}
  \Qw\equiv\left(1 -  \CD \Xw \right)\ ,   
  \eqn{Qdef}
\end{align}
and  in the last step we used the identity \cref{eq:Diracident} for a second time.

By equating the $\chi$ dependence on both sides of \cref{eq:master} we arrive at two equations.  The first requires the prefactors of the exponentials to be equal, and we will refer to this as the ``anomaly equation'':
\begin{align}
 e^{2i\CA}=\det\left( m_\omega m^{-1}\Qw\right)^{-1} = \det\left( \bar\omega \omega\Qw\right)^{-1}  \ .
  \eqn{anomgen}
\end{align}

The second equation follows from requiring that the fermion operators in the exponents must be equal,
\begin{align}
  \CD_\omega =  \Qw^{-1} \CD\ ,
  \eqn{GW1}
\end{align}
or equivalently,
\begin{align}
  \CD_\omega - \CD = \CD \Xw \CD_\omega\ ,
\end{align}
and this we call the generalized \ac{GW} equation.  If $\CD$ is invertible, the \ac{GW} equation may be written in the simple form
\begin{align}
  \omega\left(\frac{1}{\CD}-\frac{1}{ m}\right)\bar\omega =\left(\frac{1}{\CD}-\frac{1}{ m}\right) \ ,
  \eqn{GWgen}
\end{align}
which states that  the propagator is symmetric up to a constant local subtraction.  Assuming $ m$ does not couple neighboring sites, this subtraction is a delta-function in coordinate space. This relation can be further transformed to a yet simpler form by writing
\begin{align}
  \CD = m \frac{i h}{1+i h} \ ,
  \eqn{ddef}
\end{align}
in which case the \ac{GW} relation \cref{eq:GWgen} reduces to the statement that $ m h$ is invariant under the $\omega$ transformation,
\begin{align}
  \,\bar\omega (m h)  \omega  = m h\ ,
  \eqn{GWgen2}
\end{align}
or if $m$ commutes with $\bar\omega$, $h$ itself is invariant.
The requirement that $\CD$ describes a massless Dirac fermion in the continuum limit means that $\CD\to i\slashed{p} $  for  $p^2 \ll  m^2$;  thus  $h\to \slashed{p}/ m$ in that limit, which  is hermitian (assuming for now that $m$ is just a number).  If we assume that $h$ both satisfies \cref{eq:GWgen2} and is hermitian for {\it all} momenta, then we can define the unitary matrix $V = -(1-i h)/(1+i h)$ and arrive at another useful expression for $\CD$,  
 \begin{align}
  \CD = \frac{ m}{2}\left(1+V\right), \qquad  V^\dagger V = 1\ ,
  \eqn{DV}
\end{align}
with
\begin{align}
  V \to -1 + \frac{2i\slashed{p}}{ m} + O\left[\left( \frac{\slashed{p}}{ m}\right)^2\right]
  \eqn{vlimit}
\end{align}
The eigenvalues of V lie on a unit circle centered at the origin in the complex plane, and those of $\CD$ lie on a circle of radius $m/2$ centered at $m/2$. When the theory is gauged, low-lying eigenvalues of $\slashed D$ lie near $V = -1$, while large ones are mapped to the neighborhood of $V = +1$. This is familiar from the discussion in Ref.~\cite{Neuberger:1997fp}.

\subsection{Solutions to the Ginsparg-Wilson equation}
\label{sec:sols}

We now examine solutions to the \ac{GW} equation, which not only satisfy \cref{eq:GW1}, but  also satisfy $\CD \to \Dsl$ in the continuum limit $m \gg p$, in order to describe a massless Dirac fermion, and which for free fermions only vanish at zero momentum, so as to describe a single flavor in the continuum limit.  

\subsubsection{The Pauli-Villars solution}
\label{sec:PV}

Although the  \ac{GW} equation was derived in the context of a lattice regularization, it is in fact more general, and  a simple continuum solution to the \ac{GW} and anomaly equations existed decades before Ginsparg and Wilson wrote their paper: a fermion regulated by a \ac{PV} ghost. Examining this case yields insights into the nature of lattice solutions and symmetries.

We have seen that $\CD = m i h/(1+ih)$ will solve the \ac{GW} equation and describe a massless Dirac fermion in the low eigenvalue limit if $m h$ obeys  the continuum symmetries of a massless Dirac fermion, and $mh\to \slashed{p}$ for a free fermion at low $p$.  The simplest possible solution to these criteria is to simply set $i h = \slashed{D}/m$, in which case the \ac{GW} solution describes a \ac{PV} regulated fermion:
\begin{align}
  \slashed{D} \to \CD_\text{PV} = m\frac{\slashed{D}}{\slashed{D} + m} = \frac{m}{2}\left( 1 -\frac{1-\slashed{D}/m}{1+\slashed{D}/m}\right)\ ,
\eqn{DPVdef}
\end{align}
where we will take $m>0$ with the ``continuum'' limit being $m\to \infty$. The operator $\CD_\text{PV}$ is not fully regulated, but the phase of its determinant is, which is where anomalies appear.
The unitary matrix $V$ in \cref{eq:DV} is given by 
\begin{align}
   h =-i\slashed{D}/m \ ,\quad V = -\frac{1-\slashed{D}/m}{1+\slashed{D}/m} \ .
\eqn{Vpv}
\end{align}

We will show that the operator  $\Dpv$ simply illustrates two general properties of solutions to the \ac{GW} equation which we discuss below.  The first  is that the regulated $\eta$-invariant of the continuum operator -- which describes the phase of the fermion determinant -- is realized in terms of $\ln \det V$. The second is that when ghost fields are introduced to represent the PV-regulated fermion, the exact symmetry of the regulated action discovered by L\"uscher  can be simply related to the symmetry of the unregulated action.
The \ac{PV} solution will also help inform our analysis of massless Majorana fermions in \cref{sec:sol-maj}.

\subsubsection{The overlap solution}
\label{sec:OV}

The first explicit lattice solution to the \ac{GW} equation  was the overlap operator of Neuberger \cite{Neuberger:1997fp}, based on the earlier work in conjunction with Narayanan in Refs.~\cite{Narayanan:1993zzh,Narayanan:1993ss,Narayanan:1993sk} and on the domain wall fermion construction in \cite{Kaplan:1992bt}. This solution takes the $V$ matrix to be
\begin{align}
  V = \frac{\Dw}{\sqrt{\Dw^\dagger \Dw}}\ ,
  \eqn{Vov}
\end{align}
where $\Dw$ is the lattice operator for a Wilson fermion with mass $-M<0$ and Wilson coupling $r=M$ \footnote{As shown in \cite{Jansen:1992tw,Golterman:1992ub} there is actually an interesting sequence of topological phase transitions as a function of $M/r$, and taking $M/r=1$ places the theory in one of several possible topological phases.},
\begin{align}
  \Dw =  \sum_\mu  \delta_\mu\gamma_\mu-M - \frac{M }{2}\Delta\ ,
  \label{eq:Dw}
\end{align}
where $\delta_\mu$ is the covariant symmetric difference operator, and $\Delta$ is the covariant lattice Laplacian.   Without gauge fields, this gives
\begin{align}
  \Dwm(p) &=   \sum_\mu (i \gamma^\mu \sin p_\mu )+ M\left[ -1 + \sum_\mu (1-\cos p_\mu)\right] \nonumber\\
         &\to M\left(-1  + i\frac{\slashed{p}}{M} + \Order(p^2/M^2) \right)\ .
\end{align}
Evidently $V\to \left(-1 +  i\frac{\slashed{p}}{M} + \Order(p^2/M^2) \right)$ and one can see that near the corners of the Brillouin zone where doublers reside for naive lattice fermions one finds $V=+1$.  Therefore this operator behaves correctly as a massless Dirac fermion in the continuum limit.

In even spacetime dimensions, one has chiral symmetry with $\omega=\bar\omega = e^{i\alpha\gammachi}$.  Then the \ac{GW} equation as expressed by \cref{eq:GWgen2} is equivalent to $\{\gammachi,h\}=0$ or $\gammachi V \gammachi = V^\dagger$.  This latter property readily seen to be satisfied by the overlap solution.  In odd spacetime dimensions one is interested in reflection symmetry for which $\omega = -\bar\omega = \CR_1 \gamma_1$ and \cref{eq:GWgen2}  requires $\{h,\omega\}=0$, implying that $\omega V \omega^{-1} = V^\dagger$, which is also seen to be satisfied by the overlap operator.   

\subsection{An exact symmetry of the lattice action}
\label{sec:sym}

\Cref{eq:GW1} together with \cref{eq:GW0} implies that the action $\bar\chi \CD \chi$ for a \ac{GW} fermion obeys an exact L\"uscher symmetry,
\begin{align}
  \bar\chi \to \bar\chi \Qw \bar\omega  \ ,\quad \chi\to \omega\chi.
  \label{eq:trans}
\end{align}
This symmetry constrains the Feynman rules for the theory, eliminating the possibility of an additive mass renormalization for  $\chi$ in perturbation theory since  a mass term breaks the symmetry with
\begin{align}
  \bar\chi\chi \to \bar\chi \Qw\bar\omega\omega\chi\ ,
  \eqn{Lsym}
\end{align}
where $\Qw\bar\omega\omega\ne 1$  for the symmetry transformations of interest\footnote{
  This symmetry does not protect against finite nonperturbative additive mass renormalizations, such as those that can be generated by instantons as discussed in \cite{Kaplan:1986ru}.}.  
The transformation is also not a symmetry of the $\chi$ measure, with  Jacobian equal to $(1/\det \bar\omega\omega \Qw) $, which we see from the anomaly equation \cref{eq:anomgen}  exactly reproduces the $\exp(2i\CA)$ anomaly in the original continuum theory. 
This symmetry was discovered in the context  of infinitesimal chiral transformations in even spacetime dimension by L\"uscher \cite{Luscher:1998pqa,luscher1999abelian} with $\omega = \bar\omega  =1 + i \alpha\gammachi + O(\alpha^2)$, which we have generalized here to include discrete symmetries.

This symmetry may seem somewhat peculiar, but becomes transparent when considering the \ac{PV} solution.  First one simply adds a gaussian term for a spinor ghost with Bose statistics,
\begin{align}
  S_\chi \to \bar\chi \Dpv \chi + m\bar\phi \phi =m \left(\bar\chi \frac{\slashed{D}}{\slashed{D}+m}\chi + \bar\phi \phi\right),
\end{align}
integrating over the $\phi$ fields, which has no effect other than modifying the normalization of the path integral.  The fermion operator $\Dpv$ is defined in \cref{eq:DPVdef}. We then make the simultaneous change of variables
\begin{align}
  \bar\chi =\bar\chi'(1+\slashed{D}/m)\ ,\qquad \bar\phi= \bar\phi'(1+\slashed{D}/m)\ ,
\end{align}
leaving $\chi$ and $\phi$ unchanged.  Because $\bar\chi$ and $\bar\phi$ have opposite statistics, the Jacobians from these transformations cancel in the integration measure.  The action now looks like
\begin{align}
  S_\chi = \left[ \bar\chi'\slashed{D} \chi +\bar\phi' (\slashed{D} +  m)\phi\right] ,
\end{align}
which is the conventional form for \ac{PV} regularization in  perturbative applications with a massless Dirac fermion and a ghost of mass $m$. 

Using the identity
\begin{align}
  \Qw \bar\omega =  \frac{1}{(1+\slashed{D}/m)}\bar\omega (1+\slashed{D}/m)\ .
  \eqn{Qid}
\end{align}
the L\"uscher symmetry transformation of  \cref{eq:trans}  becomes very simple in terms of our new variables,
\begin{align}
    \chi \to \omega\chi\ ,\qquad \bar\chi' \to \bar\chi'\bar\omega\ ,
    \eqn{newtrans}
\end{align}
with $\phi$ and $\bar\phi'$ not transforming at all.   In other words, the transformations of  the $\chi$ and $\bar\chi'$ fields are just the symmetry transformations that leave the continuum Dirac action invariant.  Furthermore, as in the continuum,  violation of the symmetry comes from the path integral measure since \cref{eq:newtrans} has no compensating transformation of the ghost field. It is clear  that since the Feynman rules for $\chi$ and $\bar\chi'$  in this theory with ghosts respect the $\omega$ symmetry,  no  symmetry-violating operators will be generated by radiative corrections in perturbation theory.

\subsection{The anomaly equation} 
\label{sec:anomaly}

The anomaly equation \cref{eq:anomgen}  states that the continuum anomaly $\exp(2i \CA)=1/\det \Qw$ for chiral symmetry transformations (for which $\det \bar\omega\omega = 1$), while $\exp(2i \CA)=1/\det (-\Qw)$ for reflections (where $\bar\omega\omega = -1$), which in both cases equals the Jacobian for the  symmetry transformation in \cref{eq:trans}.  This relates $\CA$, which is a functional of the background  fields, to properties of the fermion spectrum.  Here we show that in even spacetime dimensions the equation reproduces the Atiyah-Singer index theorem as shown in Ref.~\cite{Luscher:1998pqa}, while in odd spacetime dimensions it reproduces the relation between the parity anomaly and the $\eta$-invariant discovered in Ref.~\cite{alvarez1985anomalies}. For recent work on the $\eta$-invariant in the context of the overlap operator, see Refs.~\cite{fukaya2020atiyah, fukaya_modtwo_2022}.

We first consider the \ac{PV} solution in both odd and even dimensions. 
The phase of the determinant for a massless Dirac fermion may be expressed as $\exp(-i\pi \eta_D(0)/2)$ , where  $\eta_D$ is defined as a regulated sum of the signs of eigenvalues of $i\slashed{D}$, and $\eta_D(0)$ is the universal value as the regulator is removed \cite{witten_anomaly_2020}.  The \ac{PV} solution to the \ac{GW} equation replaces $\Dsl$ by its regulated form $\Dpv = (m/2)(1+V)$ where $V$ is unitary.  It follows that 
\begin{align}
  \frac{\det \Dpv}{\det \Dpv^\dagger} = e^{\Tr \ln \frac{1+V\,\,}{1+V^\dagger}} = e^{\Tr \ln V}\ .
\end{align}
The eigenvalues of $V$ are $(-i \lambda/m -1)/(-i  \lambda/m+1) = -1 - 2i\lambda/m + O(1/m^2)$, and so we have
\begin{align}
  \Tr\ln V =- i\pi \sum_\lambda\frac{\lambda}{|\lambda |} + O(1/m) \equiv - i \pi \eta_D(1/m) \ .
  \eqn{pveta}
\end{align}
Thus we see that 
\beq
  \eta_D(0) = \lim_{m\to\infty} \frac{i}{\pi}\ln\det V 
\eeq
and the phase of the fermion determinant $\det \Dpv$ may be written as $ e^{-i\frac{\pi}{2} \eta_D}$. This result applies generally to solutions of the \ac{GW} equation.

In odd spacetime dimensions with a space reflection transformation as in \cref{eq:reflect} we have $\bar\omega \omega = -1$, $m_\omega = -m$  and $-\Qw =- 1+2\CD/ m = V$. Therefore the anomaly equation states that $ \CA = -\half\Tr\ln V = i\pi \eta_D/2$, correctly realizing the parity anomaly as the regulator is removed  \cite{alvarez1985anomalies}. The perturbative expansion of $\eta_D$ yields the Chern-Simons action, a result also consistent with  Ref.~\cite{CosteLuscher}.

In even spacetime dimensions for a $U(1)_A$ chiral  transformation the anomaly equation states $\exp(2i \CA)=1/\det \Qw$.  In this case it is simplest to expand to linear order in $\alpha$ and one finds
\beq
  \Qw =  1-2i\alpha/m  \CD\gammachi + O(\alpha^2)\ ,
  \eeq
 and the anomaly equation states that
\beq
 2i \CA=  \frac{2i \alpha}{ m} \Tr \gammachi \CD 
  \eqn{infqanom}
\eeq
where the continuum anomaly functional $\CA$ is proportional to $\alpha$.  The Atiyah-Singer index theorem states that the right side of the above equation should equal $-2i\alpha$ times the index of the Dirac operator, $(n_+-n_-)$, where $m_\pm$ equals the number of $\pm 1$ chirality zeromodes.  This result follows from the analysis by L\"{u}scher \cite{Luscher:1998pqa}, after taking into account the relative normalization of $a m/2$ between $\CD$ and the \ac{GW} operator analyzed in that paper. 

\subsection{Anti-linear symmetry} 
\label{sec:antilinear}

A theory that possesses an anti-linear time reversal symmetry $\psi(\bfx,t)\to \CT \psi(\bfx,-t)$ in Minkowski spacetime will respect a related anti-linear symmetry in  Euclidean spacetime that does not reverse any coordinates.   
This is simply because after replacement of $t$ with $-i\tau$, the conjugation of the $i$ in $-i\tau$ has the same effect as $t\to -t$. For this symmetry $\Omega = \bar\Omega^\dagger = \hat\CT T$ where the operator $\hat\CT$ reverses time in Minkowski spacetime but acts trivially in Euclidean, while $T$ is a unitary matrix satisfying $T^\dagger\gamma_\mu  T= \pm\gamma_\mu^T$.  When this transformation is a symmetry of the massless theory but is necessarily broken by a fermion mass term, then it will in general be anomalous and there will be corresponding \ac{GW} relations. A simple example is a massless Dirac fermion in $2+1$ dimensions where we can take the $\gamma$ matrices to be $\gamma^0 =  i\sigma_1$, $\gamma^1=\sigma_2$, $\gamma^2=\sigma_3$ and $T = \sigma_2$.  Under time reversal the fields transform as $\psi(\bfx,t)\to T \psi(\bfx,-t)$ and $\bar\psi(\bfx,t)\to \bar \psi(\bfx,-t)T$ which is a symmetry of the action for a massless Dirac fermion, but for a massive fermion the transformation flips the sign of the mass term.  In Euclidean spacetime the symmetry transformation is identical, $\psi\to T\psi$ and $\bar\psi\to \bar\psi T$,  except that there is no change in the coordinates; again one finds that the massless Dirac action is invariant but that a mass term is odd. 

Our derivation of the generalized \ac{GW} relations proceed as above, only now $\Omega$ and $\bar\Omega$ are anti-linear, while the $\omega$ and $\bar\omega$ remain as ordinary matrices. This change results in \cref{eq:GW0} being replaced by
\begin{align}
  \CD_\omega = \bar \omega \CD^* \omega\ ,\quad  m_\omega=\bar\omega  m^* \omega\ ,
  \eqn{GW0trs}
\end{align}
With these changes, the anomaly equation \cref{eq:anomgen} and the \ac{GW} equation \cref{eq:GW1} remain valid.   It is evident that $\Dpv$ satisfies this antilinear \ac{GW} equation since $h\propto\Dsl$; one can easily check that $\Dov$ satisfies it as well.

\section {Generalized Ginsparg-Wilson relations for Majorana fermions}

The edge states of topological insulators are typically massless Dirac fermions such as described in the previous section;  on the other hand, the edge states of topological superconductors  without a conserved fermion number are massless Majorana fermions.  Majorana edge states were first discussed in Ref.~\cite{Kaplan:1999jn} in the context of simulating gluinos in $d=3+1$ dimensions, and in Ref.~\cite{kitaev2001unpaired} for $d=1+1$ condensed matter systems. Here we derive the \ac{GW} relations for  Majorana fermions.

\subsection {Continuum Majorana fermions}
\label{sec:CT}

We begin by summarizing properties of continuum Majorana fermions in arbitrary $d$ dimensions, and enumerate the symmetries of interest.\footnote{For a detailed discussion of Majorana fermions in Minkowski and Euclidean spacetimes, see Ref.~\cite{stone_gamma_2021}.}

 \subsubsection{The Majorana constraint}

To obtain a single flavor of massless Majorana fermion we impose a Lorentz-covariant Majorana constraint on a massless Dirac fermion,
\begin{align}
\psi = \psi^\Ck\ ,\qquad\psi^\Ck \equiv \Ck^\dagger\bar\psi^T\ ,
\eqn{MajCon}
\end{align}
where for Lorentz invariance and self-consistency of the constraint, $\Ck$ must equal either  an {\it antisymmetric} $\CC$ matrix, or a {\it symmetric} $\CT$ matrix, $\CC$ and $\CT$ being unitary matrices which  satisfy
\begin{align}
  \Cmat \gamma_\mu \Cmat^\dagger = -(\gamma_\mu)^T \ ,\qquad
  \Tmat \gamma_\mu \Tmat^\dagger  = (\gamma_\mu)^T\ .
\eqn{c-and-t}
\end{align}
The Majorana constraint as expressed above is equally valid in Minkowski and Euclidean spacetimes.  In Ref.~\cite{stone_gamma_2021} fermions satisfying these constraints  are referred to as Majorana ($\Ck = \CC$) or pseudo-Majorana ($\Ck = \CT$); here we will refer to them as $\CC$-Majorana and $\CT$-Majorana respectively when distinguishing between them, and simply by ``Majorana'' when not.  The massless Majorana  action can then take  the form\footnote{A Majorana fermion may carry gauge charges so long as it is in a (pseudo-)real representation of the gauge group.  In that case, $\CC$ and $\CT$ will have to include the appropriate matrices to effect the similarity transformation from the generators $T_a$ to the conjugate generators $-T_a^T$.}
\beq
S = \int d^dx \,\half \psi^T \Ck \slashed{D} \psi\  .
\eeq
  Table~\ref{tab:c-and-t} lists the properties of the $\CC$ and $\CT$ matrices in different dimensions, and we see that for a single Majorana flavor we can take $\Ck = \Cmat$ in $d=2,3,4 \mod 8$, and $\Ck = \Tmat$ in $d=1,2,8 \mod 8$, while there is no solution in $d=5,6,7 \mod 8$.  Instead of one flavor, one could  consider two flavors and replace $\Ck \to \Ck\otimes \tau_2$, where $\tau_2$ is the antisymmetric Pauli matrix in flavor space. Then one requires $\Ck$ to equal either a \emph{symmetric} $\Cmat$ matrix, or an \emph{antisymmetric} $\Tmat$ matrix.  Such fermions are  sometimes referred to as symplectic Majorana fermions.  In this way one can discuss massless  fermions with a reality constraint ($\CC$-Majorana, $\CT$-Majorana, symplectic Majorana) in any dimension.  In this section we will only discuss a single flavor of massless Majorana and are therefore restricted to $d=2,3,4$. We give examples of these theories with discrete symmetry anomalies, as well as an anomalous example of symplectic Majoranas.

\begin{table}[t]
  \newcommand\Good[1]{\textbf {#1}}
   \newcommand\GoodMass[1]{\textrm {#1}}\centering
  \renewcommand{\arraystretch}{1.1}
  \begin{tabular}{c |c c c c c c c c}
    \TopRule
    $d$:        & 1 & 2 & 3 & 4 & 5 & 6 & 7 & 8\\
    \MidRule
    $\Tmat$ &  \Good{S} & \Good{S} &   $\cdot$ & \GoodMass{A} & \GoodMass{A} & \GoodMass{A} & $\cdot$ & \Good{S}  \\
    $\Cmat$ &  $\cdot$   & \GoodMass{\Good{A}} & \GoodMass{\Good{A}} & \GoodMass{\Good{A}} &  $\cdot$  & S & S &S \\
    $\gammachi$ & $\cdot$  & $-$ &$\cdot$    & $+$ &$\cdot$    & $-$ & $\cdot$  & $+$ \\
    \BotRule
  \end{tabular}
  \caption{
  The $\Cmat$ and $\Tmat$ matrices in Euclidean dimensions $d=1,\dotsc,8 \mod 8$ defined in \cref{eq:c-and-t}. S and A represent whether the corresponding matrix is symmetric or antisymmetric, while a dot indicates it does not exist. The last row denotes whether $\Cmat \gammachi \Cmat^{-1} = \Tmat \gamma^{d+1} \Tmat^{-1} = \pm (\gamma^{d+1})^T$, where $\gammachi$ is the chiral matrix for even $d$ satisfying $\{\gammachi,\gamma_\mu\} = 0$ for $\mu = 1,\ldots,d$.
  For a single Majorana flavor, only {\bf bold} entries can play the role of $\Ck$ in Majorana kinetic terms, and only antisymmetric entries (A)  can appear as $\Cm$ in Majorana mass terms.
  We refer the reader to Ref.~\cite{stone_gamma_2021} for a pedagogical discussion of this table.}
  \label{tab:c-and-t}
\end{table}

In order to follow the \ac{GW} program we must be able to define a mass term for the Majorana fermion.  This can be included in the Euclidean action as $\half\int \psi^T \mbf \psi$  where 
\beq
\mbf = \mu \Cm = -\mbf^T\ ,
\eeq
  $\mu$ being a number with dimension of mass, while $\Cm$ is required by Lorentz invariance and fermion statistics to be either an antisymmetric  $\Cmat$ or antisymmetric $\Tmat$ matrix.    No such matrix  exists in $d=1,7,8 \mod 8$.  In these cases we can consider symplectic Majoranas (two flavors) in which case $\mu$ may be replaced by $\mu\,\tau_2$ acting in flavor space, and $\Cm$ must now be a \emph{symmetric} $\Cmat$ or $\Tmat$ matrix\footnote{It is stated in Ref.~\cite{stone_gamma_2021} that $\CT$-Majorana fermions are necessarily massless, but that assumes that a mass term must have the form $\psi^T\CT \psi$.  When allowing for a $\psi^T\CC\psi$ mass term the statement is no longer true. This can be generated from a Dirac action by applying the $\CT$-Majorana constraint to a Dirac mass term of the form $i\bar\psi \gammachi \psi$.}.

 As can be seen from Table~\ref{tab:c-and-t}, the requirement that both $\Ck$ and $\Cm$ exist still restricts us to discussing $d=2,3,4$ for a single flavor.  In $d=3$ there is the unique choice $\Ck =\Cm = \CC$.  In $d=2$ we have the single choice $\Cm = \CC$ while $\Ck$ may equal $\CC$ or $\CT$.  In $d=4$, the reverse is true:  $\Ck = \CC$ while $\Cm$ may equal $\CC$ or $\CT$.  For the two mixed cases $(\CK,\CM) = (\CT,\CC)$ in $d=2$ and $(\CC,\CT)$ in $d=4$ we have $\CT$ equal to $\gammachi \CC$, up to a phase, and hermiticity in Minkowski spacetime is guaranteed if we take
 \beq
 \CM^{-1} \CK = \begin{cases} 1 & (\CC,\CC) \\ i\gammachi & (\CT,\CC),\,(\CC,\CT)\end{cases}\ .
 \eqn{mink}
 \eeq

\begin{table}[t]
  \setlength\tabcolsep{0.25em}
  \renewcommand{\arraystretch}{1.1}
  \begin{tabular}{c |c c |c |c c }
    \TopRule
 $ d$:   & \multicolumn{2}{c}{2} &  \multicolumn{1}{c}{3}&  \multicolumn{2}{c}{4} \\
    \MidRule
   $(\Ck,\Cm)$       & $\Rsym$ & $\gammachi$ & $\Rsym$ & $\Rsym$ & $e^{i\alpha\gammachi} $\\
    \MidRule
  $(\CC,\CC)$& \xmark& \xmark & \xmark & \xmark & \xmark \\
   $(\CC,\CT)$ &  $\cdot$   &$\cdot$&$\cdot$ & \chck&  \xmark \\
$(\CT,\CC)$& \chck&\xmark &   $\cdot$ & $\cdot$ & $\cdot$  \\
   $(\CT,\CT)$ &  $\cdot$   & $\cdot$   & $\cdot$   & $\cdot$   & $\cdot$    \\
      \BotRule
  \end{tabular}
  \caption{ Reflection ($\Rsym$) and chiral (discrete or continuous) symmetries for a single  massless Majorana flavor in $d=2,3,4$ for different combinations of the $\Ck$ and $\Cm$ matrices, where $\Ck$ defines the kinetic term and $\Cm$ is used as the regulating mass term.  A ``$\chck$'' indicates a non-anomalous symmetry, an ``\xmark'' denotes that the regulator choice $\Cm$ breaks the symmetry indicating a possible anomaly, and a dot means that the  $(\Ck,\Cm)$ combination does not exist. For $d\neq 2,3,4$, we need multiple flavors.
  }
  \label{tab:MajSyms}
\end{table}

 \subsubsection{Symmetries}
\label{sec:MajSym}

In dimensions $d=2,3,4$ the massless Dirac action possesses a $U(1)_V$ fermion number, reflection symmetry and charge conjugation symmetries, while in $d=2,4$  it also possesses a $U(1)_A$ chiral symmetry.  Here we examine what subgroup is left unbroken by the Majorana constraint, and then what is the effect of the regulator.

In all dimensions $U(1)_V$ fermion number symmetry is broken to a $\mathbf{Z}_2$ subgroup which acts as $(-1)^F$, an element of the Lorentz group.  What happens to the $U(1)_A$ chiral symmetry in $d=2,4$ depends on the fact that $\Ck \gammachi^T \Ck^{-1}  = -\gammachi$ in $d=2$ and $+\gammachi$ in $d=4$.
In $d=2$  in addition to $(-1)^F$ the Majorana constraint leaves unbroken a $\mathbf{Z}_2$ subgroup of $U(1)_V\times U(1)_A$ corresponding to $\psi\to \gammachi\psi$, while in $d=4$ the entire $U(1)_A$ remains unbroken.
The latter result should not be surprising since  a massless Majorana fermion in $d=4$  Minkowski spacetime is equivalent to a massless Weyl fermion, whose action possesses a $U(1)$ symmetry; this is not true in $d=2$.    

The charge conjugation symmetry of the Dirac fermion survives the Majorana constraint, but either acts trivially on the Majorana fermion, or as $(-1)^F$.
 
For reflections we consider transformations of the Dirac field $\psi(x) \to \Rsym \psi(x) = \varepsilon \gamma_1 \psi(\tilde x)$ and $\bar\psi(x) \to \Rsym \bar\psi(x) = -\varepsilon^*\bar\psi(\tilde x) \gamma_1$, where $\varepsilon$ is a phase and $\tilde x$ has the sign of $x_1$ flipped.  This is consistent with the Majorana condition \cref{eq:MajCon} if $\varepsilon =1$ when $\Ck=\CC$ and $\varepsilon = i$ when $\Ck = \CT$ and is therefore always a symmetry for the massless Majorana action.  
Note that this means that for $\CC$-Majoranas we have $\Rsym^2 = 1 $ while for $\CT$-Majoranas,  $\Rsym^2 = (-1)^F$.  

When a Majorana mass term $\mbf$ is included the  $(-1)^F$ symmetry is not broken, but the discrete chiral symmetry in $d=2$ and the continuous chiral symmetry in $d=4$ are; therefore it is reasonable to expect anomalies and \ac{GW} relations for these transformations. The situation for reflection symmetry is more complicated.   Reflection symmetry is broken by the mass term if the $\Cm$ matrix is the same as the $\Ck$ matrix, and unbroken if they are unlike (e.g. $(\Ck,\Cm) = (\CC,\CT)$ or  $(\Ck,\Cm) = (\CT,\CC)$).  Therefore we should expect reflection symmetry to be anomalous for Majorana fermions in $d=2,3$ and in $d=4$ when $\Cm = \CC$. It will not be anomalous for $\CT$-Majorana fermions in $d=2$ or $\CC$-Majorana fermions in $d=4$ with $\Cm = \CT$.  These two cases are quite different from each other, however: in $d=2$ both $\CC$- and $\CT$-Majoranas exist with only one way to regulate them (with $\CM = \CC$), and we find that reflections are anomalous in the former but not the latter.  For $d=4$ we only have a $\CC$-Majorana, but two ways to regulate, with $\CM = \CC$ or $\CM = \CT$, the former breaking reflections symmetry and the latter not.  In this case we would say that choosing $\CM = \CC$ is a poor choice of regulator, needlessly breaking the symmetry of the massless fermion, and we would not expect the symmetry to be anomalous.

We have summarized the situation with reflection and chiral symmetries in Table~\ref{tab:MajSyms}; cases for which \ac{GW} relations pertain are the entries with the ``\xmark''.

\subsection{Derivation of the relations}

Similar to the discussion of Dirac fermions in \cref{sec:gw-dirac}, we can derive a \ac{GW} relation for Majorana fermions, which we denote as $\Xi$ in the continuum.  
We follow the same block-spin prescription as for Dirac fermions and perform a transformation $\Xi\to \Omega\Xi$ which is assumed to be a symmetry of the continuum action
but not a symmetry of either the block-spin gaussian or the measure.  The analogue of \cref{eq:transformedPI} is
\beq
  e^{-\frac12 \eta^T \Dbf_{\omega}  \eta}
  &= \int d \Xi\ e^{i \Anomaly}e^{- S[ \Xi ] - (\eta - \xi)^T   \mbf_ {\omega}  ( \eta - \xi)} \ ,
\eqn{MajMaster}\eeq
where $\xi_{\bfn}$ are block-averaged lattice fields   related to $\Xi$ as in \cref{eq:discrete},
\begin{align}
\xi_\bfn = \int\! d^d\bfx\ \Xi(\bfx) f(\bfx-\bfn a)\,
\eqn{discrete-maj}
\end{align}
and $\mbf$ is an invertible, imaginary, antisymmetric matrix.  We have defined
\beq
  \Dbf_\omega = \omega^T  \Dbf\omega\ ,\qquad   \mbf_\omega = \omega^T  \mbf\omega\ ,
  \eeq
  where  $\omega$ is related to $\Omega$ in analogy with \cref{eq:omdef}, and  suppress lattice indices as before. The path integral identity we derive in  
\cref{eq:maj-iden}
allows us to recast this equation as
\begin{align}
e^{-\half \eta \Dbf_\omega\eta} = e^{i\CA} e^{\half\Tr\ln \frac{\mbf_\omega}{\mbf} \Qw} e^{-\half \eta \Qw^{-1} \Dbf \eta}\ ,
\end{align}
where 
\begin{align}
\Qw = \left( 1 - \Dbf  \Xwbf \right),\ \quad
\Xwbf = \mbf^{-1}-  \mbf_\omega^{-1} \ .
\end{align} 
Comparing both sides, we find two equations, the first of which is a generalized \ac{GW} relation for Majorana fermions
\beq
   \Dbf_{\omega} = \Qw^{-1} \Dbf \ .
\eqn{GWMaj}\eeq
This can be rewritten in a form analogous to the conventional \ac{GW} relation as
\begin{align}
  \Dbf_\omega - \Dbf = \Dbf \Xwbf \Dbf_\omega .
\eqn{hsgw}
\end{align}
      If there are no zeromodes, then $\Dbf$ is invertible and the \ac{GW} equation is equivalent to
\begin{align}
\omega^T\left(\frac{1}{\Dbf}-\frac{1}{\mbf}\right) \omega = \left(\frac{1}{\Dbf}-\frac{1}{\mbf}\right)\ ,
\eqn{GWMaj2}
\end{align}
similar to what we found for the Dirac case in \cref{eq:GWgen}.

As in the Dirac case, the second equation obtained is the anomaly equation, 
\begin{align}
    e^{i\CA}  =\frac{1}{\sqrt{\det\frac{\mbf_\omega}{\mbf}  \Qw}}.   
    \eqn{MajAnomEq}
\end{align}
As we shall show, the square root is well defined.

\subsection{Solutions to the Majorana Ginsparg-Wilson equation}
\label{sec:sol-maj}

Just as we identified both the \ac{PV} and overlap solutions to the \ac{GW} relations for Dirac fermions, we can do the same for Majoranas.  The \ac{PV} solution allows one to easily derive certain useful  properties of a solution which generalize.

\subsubsection{Pauli-Villars solution}
\label{sec:pv-majorana}

If we write
\begin{align}
\Dbf = \mbf \frac{ih}{ih+1}
\eqn{Dbfhform}
\end{align}
then the \ac{GW} relation in \cref{eq:GWMaj2} is equivalent to the statement
\begin{align}
\omega^T \mbf h \omega = \mbf h\ ,
\eqn{GWMaj3}
\end{align}
or that $\mbf h$ possesses the same symmetry as the continuum operator for a massless Majorana fermion, $\CK\slashed{D}$.  Furthermore, the continuum limit  requiring that $\Dbf \to i\Ck \slashed{p}$ in the low momentum limit for a free fermion implies that
$h\to \mbf^{-1} \CK \slashed{p}$.  As in the Dirac example discussed in \cref{sec:PV}, the simplest solution to simply set $\mbf h = \CK\slashed{D}$, and the interpretation to this solution of the \ac{GW} equation is a \ac{PV} regulated Majorana fermion,
\beq
\Dbfpv = \mu \Ck\slashed{D} \frac{1}{  \Cm^{-1}\Ck \slashed{D}  + \mu  }\ , 
\eqn{pv-majorana}
\eeq
where   $ \Cm^{-1}\Ck = 1$ or $ \Cm^{-1}\Ck=\pm i\gammachi$, depending on which of the ``\xmark'' cases in Table~\ref{tab:MajSyms} one is discussing,  while $\mu$ is the \ac{PV} mass scale. Given that $\Ck \slashed{D} $ and $\Cm$ are antisymmetric, it is easy to show that $\Dbfpv $ is antisymmetric as well.

This solution can be written as
\begin{align}
\Dbfpv = \frac{ \mbf}{2}\left(1 +  \Vmaj\right)\ ,\quad  \Vmaj =- \frac{  \mu-\Cm^{-1}\Ck \slashed{D}  }{ \mu+ \Cm^{-1}\Ck \slashed{D}  }\ ,
\eqn{DpvVform}
\end{align}
where  $ \Vmaj$ is a unitary matrix.  The eigenvalues of $ \Vmaj$ lie on a circle, as in the Dirac case, where zeromodes of   $\slashed{D}$ are mapped to $\Vmaj=-1$, while infinite eigenvalues  are mapped to $\Vmaj=+1$. 
For the cases where $\CM = \CK = \CC$, $ \Vmaj$ is the same matrix we found for Dirac \ac{PV} solution, \cref{eq:Vpv}.

Various general properties of $\Vmaj$ can be derived from the expression in \cref{eq:DpvVform}.  Antisymmetry of $\Dbfpv$ implies that
\begin{align}
\mbf \Vmaj \mbf^{-1} = \Cm \Vmaj\Cm^{-1} = \Vmaj^T
\label{eq:Vmsym}
\end{align}
Since $\Vmaj$ is unitary, we can its eigenvalue equation as  $\Vmaj \psi_n = e^{i\theta_n}\psi_n$, while it follows from  \cref{eq:Vmsym} that $\Vmaj \Cm^\dagger\psi^*_n = e^{i\theta_n}\Cm^\dagger \psi^*_n$.  Furthermore, $\psi_n$ and $\CM^\dagger\psi_n^*$ are mutually orthogonal due to the antisymmetry of $\CM$.  Therefore it follows that the eigenvalues of $\Vmaj$ are all doubly degenerate.  This will be relevant below when we discuss the square root of the determinant  of $\Vmaj$.

Next we show how symmetries impact the eigenvalue spectrum of $\Vmaj$. In the continuum,  reflection symmetry for a Dirac fermion takes $\psi \to (\gamma_1\CR_1)\psi$ where $\CR_1$ reflects the $x_1$ coordinate, with $(\gamma_1 \CR_1)\slashed{D}(A) (\gamma_1 \CR_1) = -\slashed{D}(\tilde A)$, assuming that background fields $A$ are also suitably reflected to $\tilde A$.  It follows that since $ \Cm^{-1}\Ck$ equals one in the $(\CC,\CC)$ theories and $i\gammachi$ in the $(\CC,\CT)$ and $(\CT,\CC)$ theories that
\beq
  (\gamma_1\CR_1) \Vmaj  (\gamma_1\CR_1) = \begin{cases} \Vmaj^\dagger & (\CC,\CC) \\ \Vmaj & (\CC,\CT),\,(\CT,\CC) \end{cases}\ ,
 \eqn{Vmref}
  \eeq
again assuming a reflection of background fields in the $\Vmaj$ matrices on the right.  

The effect of $\gammachi$ in $d=2,4$ is seen to be the same as seen in the Dirac case, namely
\begin{align}
\gammachi \Vmaj\gammachi = \Vmaj^\dagger\ .
\eqn{Vmchi}
\end{align}

We will be interested in the anomalous symmetries marked by the ``\xmark'' in Table~\ref{tab:MajSyms}.   We see that in each of these cases we have a unitary matrix $\CU$  satisfying $\CU \Vmaj \CU^\dagger =   \Vmaj^\dagger$. 
 This implies that if $\Vmaj \psi_n = e^{i\theta_n}\psi_n$, then $\Vmaj  \CU^\dagger\psi_n = e^{-i\theta_n}\CU^\dagger\psi_n$, and therefore, not only are all eigenvalues of $\Vmaj$ doubly degenerate, but the $V \neq \pm 1$ eigenvalues also come in complex conjugate pairs\footnote{One can relax the assumption that $\Vmaj$ is unitary and still conclude the eigenvalues come in $\{\lambda, \lambda^{-1}\}$ pairs for $\lambda \neq \pm 1 $.}.

\subsubsection{Overlap solution}

Armed with insight from the above \ac{PV} solution, it is straightforward to  find a lattice overlap solution to  the Majorana \ac{GW} equation,
\begin{align}
  \Dbfov &= \frac{\mbf}{2} (1 + \Vmaj) \\
  \Vmaj &= \frac{\Dw}{\sqrt{\Dw^\dagger \Dw}}
\end{align}
where
\begin{align}
  \Dw &= \Cm^{-1} \Ck \gamma^\mu {\delta_\mu} -\mu (1 + \Delta/2),
  \label{eq:Dw-maj}
\end{align}
where $\delta_\mu$ and $\Delta$ are the lattice derivative and Laplacian respectively.  The overlap solution for $\Vmaj$ obeys the properties we found for the \ac{PV} solution, \crefrange{eq:Vmsym}{eq:Vmchi}.  Without gauge fields and in momentum space,
\begin{align}
  \Dwm(p) &=  \Cm^{-1} \Ck \sum_{\mu} \gamma^\mu i \sin(p_\mu)  
 \nonumber\\
  &\quad +  \mu\left[-1 + \tsum_{\mu} (1 - \cos(p_\mu))\right].
\end{align}
Near the origin $p\ll \pi/a$ we have
\begin{align}
  \Dwm(p) &= \Cm^{-1} \Ck i \slashed{p}  + \Order(p^2/\mu^2).
\end{align}
and thus
\beq
  \Vmaj &=&-1+  \frac{ \Cm^{-1} \Ck i \slashed{p}}{|\mu|}  + \Order(p^2/\mu^2),\cr
  \Dbfov(p) &=& \frac{\mbf}{2}(1 +  \Vmaj )  = \frac{i}{2}\Ck   \slashed{p}   + \Order(p^2/\mu^2),
\eeq
 the correct continuum dispersion relation for a massless Majorana fermion.  At the corners of the Brillouin zone, however, $\mu\left[-1 + \tsum_{\mu} (1 - \cos(p_\mu))\right]>0$ and $\Vmaj \simeq 1$ so that $ \Dbfov$ does not have low-lying  eigenvalues associated with these states.

\subsection{Exact lattice symmetry for Majorana fermions}

As in the Dirac case for the anomalous chiral and parity symmetries, the Majorana \ac{GW} action respects exact versions of the various anomalous symmetries listed in \cref{tab:MajSyms}, with the Jacobians of the transformations reproducing the anomaly $\CA$. Here we discuss the exact form  respected by the \ac{GW} operator for each of the symmetries listed in that table.  In the next subsection we examine the anomaly equation \cref{eq:MajAnomEq} and show how the Jacobians of the exact lattice symmetry transformations correctly reproduce the known continuum anomaly $\CA$.

 The Majorana \ac{GW} equation in \cref{eq:hsgw} implies an exact L\"uscher symmetry for any antisymmetric $\Dbf$ which satisfies it.
 To see this, we can rearrange the Majorana \ac{GW} relation as
\begin{align} 
  \Dbf=
\sqrt{\Qw}\ 
  \Dbf_\omega \ 
\sqrt{\Qw}^T
  \label{eq:gw-maj-Qw}
\end{align}
where $\Qw = (1 -  \Dbf \Xwbf)$  and  $\Qw^T = (1 -   \Xwbf\Dbf)$. 

Care must be taken in the definition of the square root. Our convention is to define the square root of $\Qw$ to be the unique matrix with the same eigenvectors as $\Qw$ and whose eigenvalues are the square roots of the eigenvalues of $\Qw$ with non-negative real part.  We take the cut for the square root to be along the negative real axis, and for negative real eigenvalues of $\Qw$ we will either define the corresponding eigenvalues of $\sqrt{\Qw}$ to all lie on the positive imaginary or negative imaginary axes, denoting the choice by $\sqrt[\pm]{\Qw}$ respectively.  We will see in \cref{sec:U1A} that both choices come into play. When giving general arguments we will omit the $\pm$ designation.

\Cref{eq:gw-maj-Qw} can be derived by noting that $\Qw \Dbf  =\Dbf \Qw^T$, and so $\sqrt{\Qw} \Dbf = \Dbf \sqrt{\Qw}^T$. For a discrete symmetry transformation, $\Qw = \sqrt{-\Vmaj^T}$. 
Therefore, corresponding to the continuum symmetry $\eta \to \omega \eta$, any \ac{GW} regulated lattice action has an exact L\"uscher symmetry
\begin{align}
  \eta \to \omega \sqrt{\Qw}^T \eta \ .
  \label{eq:maj-sym-exact}
\end{align}
In terms of $\Dbf = \frac{\mbf}{2}(1 + \Vmaj)$, we can write 
\begin{align}
  \sqrt{\Qw}^T&= \left[1 - \Xwbf \Dbf \right]^{1/2} \cr
  &= \left[\frac12 (1 + \Deltaw )  - \frac12 (1 - \Deltaw) \Vmaj \right]^{1/2}.
  \label{eq:qwt}
\end{align}
The low-energy  ($\mbf \to \infty$) limit we have $\Xw \to 0$ and  $\Qw \to 1$. The symmetry transformation then reduces to $\eta \to \omega \eta$, as would be expected in the continuum limit.

Although the action is invariant under this symmetry, the fermion measure is, in general, not. The transformation in \cref{eq:maj-sym-exact} produces a Jacobian $\det(\omega \sqrt{\Qw}^T)$.  We will see in next subsection that this Jacobian reproduces the correct anomaly, once care is taken with eigenvalues of $\Qw$ which lie on the cut of the square root. While the exact symmetry in \cref{eq:maj-sym-exact} is completely general for any (continuous or discrete) symmetry, we will restrict now to the symmetries discussed in \cref{tab:MajSyms} for a single-flavor Majorana.  We will also assume $\Vmaj$ is unitary for simplicity, and obeys the properties in \crefrange{eq:Vmsym}{eq:Vmchi}, but the arguments can be generalized for the non-unitary case. 

\subsubsection{\texorpdfstring{Discrete chiral and reflection $\ZZ_2$ symmetries in $d=2,3$}{Discrete chiral and reflection Z2 symmetries in d=2,3}}

In $d=2,3$ a massless $\Cmat$-Majorana has a $\ZZ_2$ reflection symmetry which is anomalously broken by the regulating mass term. The same is true in $d=2$ for the discrete chiral symmetry for either type of  Majorana.

In all these cases of a $\ZZ_2$ symmetry broken by the regulator, the mass term flips sign, $\mbf_\omega \mbf^{-1} = -1 $. In this case $\Qw^T= -\Vmaj$ and  the exact symmetry takes the simple form
\begin{align}
  \eta \to \omega \sqrt{-\Vmaj} \eta.
\end{align}
where $\omega = \CR_{1} \gamma_1$ for the reflection symmetry and  $\omega = \gammachi$ for the discrete chiral symmetry. We can equally well define the square root as either $\sqrt[\pm]{-\Vmaj}$ for these discrete symmetries.
We will analyze the Jacobian in the next subsection and compare with the continuum anomaly.

The massless $\Cmat$-Majorana in $d=4$ and the $\Tmat$-Majorana in $d=2$ also have reflection symmetries $\Rsym$, but they are nonanomalous since a regulating mass term exists which is $\Rsym$-invariant. In such cases, a \ac{GW} formulation is trivially invariant under the corresponding continuum symmetry, without any modification.

\subsubsection{\texorpdfstring{$U(1)_A$ symmetry in $d=4$}{U(1)A symmetry in d=4}}
\label{sec:U1A}

In $d=4$, the continuum $\Cmat$-Majorana fermion has an anomalous continuous $U(1)_{A}$ symmetry $\eta \to e^{i \alpha \gammachi} \eta$, since either choice of the regulating mass term breaks this symmetry, as discussed in \cref{tab:MajSyms}. Under the $U(1)_A$ transformation $\omega = e^{i \alpha \gammachi}$, the mass term transforms such that $\mbf_\omega^{-1} \mbf = e^{-2 i \alpha \gammachi}$. The exact lattice symmetry of \cref{eq:maj-sym-exact} can then be simplified to 
\begin{align}
  \eta \to e^{ i \alpha \gammachi /2 } \left\{ \cos\alpha  - i \gammachi \Vmaj \sin\alpha \right\}^{1/2} \eta.
  \label{eq:majorana-u1}
\end{align}
In the low-energy limit, $\Vmaj \to -1$, and this reduces to the continuum symmetry, $\eta\to e^{ i \alpha \gammachi}\eta$.

This continuum $U(1)_A$ for Majorana fermions descends from the anomalous $U(1)_{A}$ symmetry for Dirac fermions upon imposing a reality condition.
However, the Majorana $U(1)_A$ symmetry in \cref{eq:majorana-u1} is distinct from the Dirac case of \cref{eq:trans}.
So one might wonder how these two definitions of the symmetry are related.
To reconcile this, we note that for Majorana fermions, a straightforward analogy of \cref{eq:trans} is not possible, since for Dirac fermions we exploited the freedom to transform  $\bar\psi$ and $\psi$ independently,  which is not consistent with the Majorana constraint.  However, that choice for how the Dirac fields transform was not unique.
To illustrate this, we consider the same example considered in Ref.~\cite{Luscher:1998pqa}, a Dirac fermion in $d=4$ with $\CD = \frac{m}{2}(1+V)$, only assuming that $\CD$ obeys the \ac{GW} equation so that $\gammachi V \gammachi = V^{-1}$. The infinitesimal  transformation corresponding to \cref{eq:trans} is
\begin{align}
 \delta\chi = \gammachi\chi ,\qquad \delta\bar\chi= \bar\chi (-V \gammachi)\ , 
\end{align}
where in the continuum limit ($V\to -1$) this reduces to the conventional chiral symmetry transformation.
However the action $\frac{m}{2}\int\bar\chi(1+V)\chi$ is invariant under the more general transformation, namely
\begin{align}
 \delta\chi = \gammachi f(V)\chi ,\qquad \delta\bar\chi= \bar\chi g(V) \gammachi\ ,\qquad 
 \label{eq:fgtrans}
\end{align}
with $f(-1)=g(-1)=1$, provided that the functions $f,g$ satisfy
\begin{align}
 g(V) V^{-1} = \gammachi f(V) \gammachi
  \label{eq:fg}
\end{align}
projected on the subspace orthogonal to $V=-1$.
Furthermore, one finds that so long as \cref{eq:fg} is satisfied, the Jacobian of the transformation reproduces the correct anomaly.  \Cref{eq:trans} satisfies this with $f=1$ and $g=-V$; alternatively, a symmetric form compatible with Minkowski spacetime where $\chi$ and $\bar\chi$ are not independent is $f =g = (1-V)/2$ \cite{Luscher:1998pqa,Creutz:2001wp}.
It is easily checked that this infinitesimal transformation keeps the Majorana action invariant.
This result holds equally well for both $(\CC,\CC)$ and $(\CC,\CT)$ regularizations.
However, a drawback with this transformation is that $\gammachi(1-V)/2$ does not generate a compact $U(1)$ symmetry, its eigenvalues not in general being integer. 
 
\Cref{eq:fg} suggests a different symmetric form consistent with the Majorana constraint, however:  $f=g=\sqrt{-V}$, which is precisely \cref{eq:maj-sym-exact}.
This choice has the feature that $\gammachi\sqrt{-V}$ is hermitian and has $\pm 1$ eigenvalues so that it generates a compact $U(1)$ symmetry; on the other hand, one must take care of the branch cut of the square root, as discussed following \cref{eq:gw-maj-Qw}, where we defined $\sqrt[\pm]{-V}$ as $\pm i$ when acting on the eigenstate of $V$ with  eigenvalue $V=1$ which lies on the cut for the square root.  Such eigenvalues correspond to the corners of the Brillouin zone for the overlap solution, or infinite momentum for the \ac{PV} solution. The  solution to \cref{eq:fg} is then $f = \sqvpM$ and $g=\sqvmM$ (or with the $\pm$ reversed). However, this is still not a satisfactory symmetry for the $d=4$ Majorana fermion because the different treatment of the branch cut for $\psi$ and $\bar\psi$ is not consistent with the Majorana constraint, $\psi = \CC^\dagger \bar\psi^T$.  

We are forced then to define the  ``pseudo-L\"uscher symmetry'' with $f=g = \sqvpM$ which is consistent with the Majorana constraint, but fails to be a symmetry of the action for $\Vmaj=+1$ eigenstates.  This is a failure at short distance and does not destroy the desirable feature of L\"uscher symmetry that chiral symmetry violating operators can only be multiplicatively renormalized.  One does lose the feature that the Jacobian of the transformation reproduces the correct anomaly, as there now appears a spurious contribution $2(\bar n_+-\bar n_-)$  where $\bar n_\pm$ are the number of $\pm$ chirality $\Vmaj=1$ modes, but this is exactly compensated by a symmetry violation in the action under such a transformation. Typically, the chiral anomaly comes from a transformation under which the action is invariant but the measure is not, so that the path integral acquires a phase under a transformation which is classically a symmetry. Although this symmetry is violated in $V = 1$ subspace, the path integral acquires the same phase under such a transformation as it would choosing $f = \sqrt[+]{-\Vmaj}$ and $g = \sqrt[-]{-\Vmaj}$. Integrating over such modes one recovers the expected anomalous Ward-Takahashi identities, so that this symmetry has the same properties as any anomalous quantum symmetry.

If gauge fields or other parameters in the theory are varied  such that an eigenvalue of $\Vmaj$ passes through $+1$, the operator $\sqvpM$ will be discontinuous.  Because of this nonanalyticity, our $U(1)_A$  transformation is nonlocal in spacetime, thereby evading a recent no-go theorem \cite{fidkowski2023no}. As  we showed at the end of \cref{sec:pv-majorana}, however, the eigenvalues of $\Vmaj$ are doubly degenerate, and therefore the determinant of $\sqvpM$ is continuous at such points.

\subsection{The anomaly equation}

We have seen in \cref{eq:MajAnomEq} that the anomaly equation gives
$
 e^{i\CA}  =(\det\frac{\mbf_\omega}{\mbf}  \Qw)^{-1/2}$.   
On the other hand, the exact symmetry of the \ac{GW} operator is not symmetry of the path integral measure and gives rise to a Jacobian
$1/\det(\omega \sqrt{\Qw}^T)$.  The first thing we will show is that these are equivalent.
Note that the square of the anomaly from \cref{eq:MajAnomEq}  is clearly equal to the square of the Jacobian, so these two agree up to a sign. It is easy to see that the anomaly equation and the Jacobian  agree for any infinitesimal symmetry transformation, and so it is only the case of  discrete symmetries that needs careful examination.

For the anomalous discrete symmetries in \cref{tab:MajSyms} we have $\mbf_\omega\mbf^{-1} = -1$ and so \cref{eq:qwt} gives us $\Qw^T = 
-\Vmaj$.  The matrix $\Vmaj$ has eigenvalues $-e^{i\theta_n}$ with $-\pi < \theta_n \le \pi$, where the $\theta_n$ are doubly degenerate and which occur in $\pm$ pairs for $\theta_n\ne 0,\pi$ (due to reflection and chiral symmetry in odd and even dimensions, respectively). Thus there we can write
\newcommand\nV{\nu}
\begin{align}
\text{dim }  \Vmaj = \nV_+ + \nV_- + \nV_c\ ,
\eqn{Vdim}
\end{align}
where $\nV_\pm$ are the numbers of eigenvalues of $\Vmaj$ equal to $\pm 1$ and $\nV_c$ is the number of complex eigenvalues (the $\pm$ is not related to chirality). Here, $\nV_\pm$ are even integers and $\nV_c$ is a multiple of 4. The  eigenvalues of $\sqrt[+]{-\Vmaj}$ are then  $e^{i\theta_n/2}$  and only the $\theta_n = \pi$ eigenvalues contribute nontrivially to its determinant, so that  $\det{ \sqrt[+]{\Qw}^T} = i^{\nV_{+}} = (-1)^{\nV_+/2}$. Since $\nu_+$ is even and $i^{\nu_+} = (-i)^{\nu_+}$, it makes no difference which of the two definitions of the square root $\sqrt[\pm]{-\Vmaj}$ is used. The matrix $\omega$ is traceless and squares to 1, so $\det \omega = (-1)^{\text{dim }\Vmaj/2}$.  Thus we get
\begin{align}
\det(\omega \sqrt{\Qw}^T) = (-1)^{\text{dim }\Vmaj/2}(-1)^{\nV_{+}/2} = (-1)^{\nV_{-}/2}\ ,
\end{align}
where we used \cref{eq:Vdim}.
Since $\nV_-$ corresponds to the zeromodes of $\Dbf$, we find that the Jacobian of our exact symmetry yields the mod 2 index of $\Dbf$.
In comparison, for our anomaly equation in \cref{eq:MajAnomEq}  we compute $\sqrt{\det \mbf_\omega\mbf^{-1}  \Qw} =\sqrt{\det \Vmaj}$ which directly gives the same result, $(-1)^{\nV_{-}/2}$, since only the $-1$ eigenvalues of $\Vmaj$ contribute.

\subsection{Examples}

In this section, we present examples of the Majorana anomaly equation in which the \ac{GW} construction reproduces global anomalies of Majorana fermions.
In all the examples below we have $\mbf_\omega^{-1}\mbf = -1$ and $\Xwbf = 2 \mu^{-1} \Cm$, so the specification of $(\Ck,\Cm)$ matrices completely fixes the \ac{GW} equation and its solutions.

\subsubsection{Two dimensions}
In two dimensions it is possible to have either  a 2-component $\CC$- or  $\CT$-Majorana fermion, but only $\Cmat$ can be chosen as the mass term in the regulator. In this section, we show that the \ac{GW} formulation reproduces known nonperturbative anomalies for both these theories.

The continuum $\Tmat$-Majorana theory with the action $\int \eta^T \Tmat \slashed{D} \eta + m \eta^T \Cmat \eta$ corresponds to the field theory of the Kitaev chain. This has an exact (non-anomalous) reflection symmetry with $\Rsym^2 = (-1)^F$ (equivalent to $\Tsym^2 = 1$ in Minkowski space), but the mass term breaks a discrete chiral symmetry: $\eta \to \gammachi \eta $,  suggesting an anomaly for the discrete chiral symmetry.
Indeed, the anomaly is given by the mod-2 index of the Dirac operator on modes of one chirality \cite{kapustin_fermionic_2015,witten_fermion_2016,karch_web_2019a}.
With the choice $(\Ck, \Cm) = (\Tmat, \Cmat)$, we can formulate \ac{GW} equation for the massless $\Tmat$-Majorana fermion and solutions to it.
The exact L\"uscher symmetry corresponds to $\eta \to \gammachi \sqrt{-\Vmaj} $.
As shown in the previous section, the Jacobian gives $\det \omega \sqrt{-\Vmaj} = (-1)^{\nu_-/2} $, where $\nu_-$ is the number of modes with $\Vmaj=-1$, which correspond to exact zeromodes of $\Dbf$.   We have seen in \cref{sec:pv-majorana} that $\gammachi\Vmaj\gammachi = \Vmaj^\dagger$, so the $\Vmaj=-1$ eigenmodes can be taken to be simultaneous eigenstates of $\gammachi $.   We also showed that the eigenvalues of $\Vmaj$ are doubly degenerate with eigenfunctions $\psi$ and $\Cm^\dagger  \psi^*$.   The   $d=2$ relation $\CM  \gammachi \CM^{-1} = -  \gammachi^T$ then tells use that the eigenvalues of the $\Vmaj=-1$ eigenmodes come in $\pm$ chiral pairs.   Thus we can write $\nu_- = n_+ + n_{-} = 2 n_+$, where $n_\pm$ are the number of positive and negative chirality zero modes of $\Dbf$.  Therefore, the Jacobian of the discrete chiral L\"uscher symmetry reduces to $(-1)^{n+}$, which is precisely the continuum result. On a torus with periodic boundary conditions in both directions, $n_+ = n_- = 1$, and therefore we find a nontrivial anomaly.
 
Next we consider the case of a single  $\Cmat$-Majorana fermion in $d=2$.  This theory has a reflection symmetry $\Rsym \eta(x) = \gamma_1\eta(\tilde x)$ with $\Rsym^2 = 1$ and a discrete chiral symmetry, but the $\Cmat$ mass term violates them both. It is known that  this theory has a mixed anomaly between $\Rsym$ and $(-1)^F$ symmetry which can be detected in the continuum by computing a mod-2 index on a two-dimensional unorientable manifold
\cite{witten_fermion_2016}.
In the \ac{GW} formulation 
defined with $(\Ck,\Cm) = (\Cmat, \Cmat)$,
this can again be obtained simply from the Jacobian of the exact reflection symmetry for the \ac{GW} Majorana fermion.
The L\"uscher symmetry is $\eta \to \gamma_1 \sqrt{-\Vmaj} \eta(\tilde x)$.
By the same argument as before, the Jacobian for this symmetry reduces to $(-1)^{\nu_-/2}$.
On a torus with periodic boundary conditions, we have two zero modes. Then $(-1)^{\nu_-/2} = -1$ and therefore the measure acquires a sign under the reflection symmetry.

\subsubsection{One dimension}

In one dimension, fermi statistics forbid any mass term for a $N=1$ flavor 1-component Majorana,  To apply the \ac{GW} construction, we therefore need at least $N=2$ flavors, which allows for the choice $(\Ck, \Cm) = (1, \tau_2)$ with the continuum action $S = \int \eta^T \del_0 \eta + \mu\ \eta^T \tau_2 \eta$, where $\eta^T = (\eta_1,\, \eta_2)$ and $\eta_{1,2}$ are one-component Majoranas. Note that the kinetic term is invariant under a $\Rsym^2 = (-1)^F$ reflection symmetry which acts as $\Rsym \eta(t) = i \eta(-t) $, but the mass term is odd under this symmetry.  Indeed, this system corresponds to the edge modes of the Fidkowski-Kitaev chain and is afflicted by a well-known $\ZZ_8$ anomaly between $\Rsym$ and $(-1)^F$ \cite{Fidkowski:2009dba,witten_fermion_2016}.

With $(\Ck, \Cm) = (1, \tau_2)$, we can proceed with the \ac{GW} construction for $N=2$ flavors.
If $n_{a}$ is the number of zero modes corresponding to flavor $a$, the antisymmetric mass matrix $\Cm = \tau_2$ ensures a doubling of spectrum and $n_1 = n_2$.
As before, the Jacobian for the exact reflection symmetry produces a phase of $(-1)^{\nu_{-}/2}$ and $\nu_{-} = 2 n_{1}$. Since $n_1 = n_2 = 1$ on a circle with periodic boundary conditions, this represents an anomaly. It is interesting to note that since for two flavors we find a $\ZZ_2$ anomaly, the \ac{GW} formulation implies a $\ZZ_4$ anomaly for a single Majorana flavor, even though a mass term cannot be written in such a theory.   The correct answer though is that there should be a $\ZZ_8$ anomaly.  See a discussion  in Ref.~\cite{kaidi2020topological}, eq. (2.26), which suggests that the $\ZZ_4$ follows from being insensitive to a bosonic anomaly. 

\section{Conclusions}

The early work on anomaly descent equations \cite{atiyah1984dirac,zumino1985chiral,stora1984algebraic} and their embodiment in the bulk/boundary correspondence of gapped fermions \cite{Callan:1984sa} has been greatly expanded upon in recent years with the discussions about more general classes of topological materials and a wider variety of anomalies (see, for example, \cite{witten_anomaly_2020}).  A parallel development from lattice gauge theory had shown that for the case where the boundary theory is described by a Dirac fermion, one can describe the physics, including chiral anomalies, in terms of a theory that makes no reference to the bulk.  Such a theory is governed by the Ginsparg-Wilson equation \cite{Ginsparg:1981bj} which has an explicit solution in the form of the overlap operator \cite{Neuberger:1997fp}. In this paper we have shown how to generalize  the \ac{GW} analysis to encompass a wide  range of topological materials that have been classified in the condensed matter literature, focusing on topological superconductors with Majorana edge states, which are less familiar to those working in lattice gauge theory.     In each case we have generalized the notion of a L\"uscher symmetry: an exact symmetry of the lattice action which becomes identical to the continuum symmetry in the continuum limit, under which the the lattice integration measure transforms by the appropriate phase to account for the anomaly.  The class of theories for which we can derive  \ac{GW} relations contain only those for which a fermion mass term can be included, and therefore does not include chiral gauge theories, for example.

Open questions remain.  In particular the Dai-Freed anomalies discussed in the literature \cite{dai1994eta,witten_fermion_2016,witten2016parity} do not seem apparent in this approach. Thus, for example, one of the results in this work was the derivation of a $\ZZ_4$ discrete time reversal anomaly for the Fidkowski-Kitaev Majorana chain, but not the full $\ZZ_8$ anomaly known to be correct \cite{witten2016parity}.   On the other hand, we know that the overlap operator which solves the \ac{GW} equation is derived by integrating out bulk modes from a higher dimension theory \cite{Kaplan:1992bt,neuberger1998vectorlike}, which one would expect ``knows'' about such anomalies.

The solutions presented here for the generalized \ac{GW} are all formulated in Euclidean spacetime, and  are not amenable to a Hamiltonian description of the physics in continuous time.  Furthermore, not being ultra-local in Euclidean time makes the derivation of a transfer matrix and Hamiltonian problematic.  We note, though, that we defined the anomalous $U(1)_A$ pseudo-L\"uscher symmetry  that acts on $\psi$ and $\bar\psi$ in a way consistent with a Minkowski interpretation, and find  that it is not analytic in momentum, and hence not a local operator in spacetime, evading the no-go theorem in Ref.~\cite{fidkowski2023no}.  Pursuing a Hamiltonian formulation of the ideas presented here in order to render the results more applicable to real condensed matter systems seems like another avenue to explore in the future.

Finally, while it has been assumed that the fermions we consider are propagating in smooth, background gauge and gravitational fields, we have not examined in any detail the role played by the role played by unorientable manifolds, which are understood to play an important role in understanding the reflection (time reversal) anomalies \cite{witten2016parity}.

\section{Acknowledgements}
We wish to thank L. Fidkowski and J. Kaidi for useful comments.
HS would like to thank Hanqing Liu, Mendel Nguyen and Yi-Zhuang You for helpful conversations.
This research is supported in part by DOE Grant No. DE-FG02-00ER41132.
HS is also supported
by the Department of Energy through the Fermilab QuantiSED program in the area of ``Intersections of QIS and Theoretical Particle Physics." 
Fermilab is operated by Fermi Research Alliance, LLC under contract number DE-AC02-07CH11359 with the United States Department of Energy.
HS was also supported in part by the U.S. Department of Energy, Office of Science, Office of Nuclear Physics, InQubator for Quantum Simulation (IQuS) under Award No. DOE (NP) DE-SC0020970. DBK acknowledges the hospitality of both the  CCPP at NYU  and the IFT at UAM, Madrid, where parts of this work were completed. HS acknowledges the hospitality of 
Aspen Center for Physics, which is supported by National Science Foundation grant PHY-2210452; his participation  was supported in part by the Simons Foundation.

\appendix

\section{Derivation of path integral identities}
\label{sec:appendix_identities}

Here we derive two identities used in this paper.  For Dirac fermions and an invertible hermitian operator $A$ we write
\begin{align}
e^{-\bar\chi A \chi} = \det A\, \int d\psi d\bar\psi \, e^{\bar\psi A^{-1} \psi +\bar\psi \chi + \bar\chi \psi}\ .
\end{align}
It follows  that
\begin{align}
  e^{\partial_\chi B \partial_{\bar\chi}}\,e^{-\bar\chi A \chi}
  &=  \det A\,  \int d\psi d\bar\psi \, e^{-\bar\psi \left(A^{-1} -B\right)\psi + \bar\psi \chi + \bar\chi \psi} \nonumber\\
 &= \det\left(1-AB\right) \,e^{-\bar\chi\left( \frac{1}{1-AB} A \right)\chi}\nonumber\\
 &= e^{\Tr\log(1-AB)} \,e^{-\bar\chi\left( \frac{1}{1-AB} A \right)\chi}\ .
 \eqn{Diracident}
 \end{align}
The above result extends to non-invertible $A$.
 
 An analogous identity can be derived for Majorana fermions.   Assuming an invertible imaginary  antisymmetric operator $\Abf$ we have
 \begin{align}
 e^{\half  \eta \Abf  \eta} = \frac{1}{{\rm Pf}\left( \Abf^{-1}\right)}\, \int d \nu \, e^{\half  \nu \Abf^{-1}  \nu + \nu   \eta }\ .  
 \end{align}
From this one derives for antisymmetric $\Bbf$
\begin{align}
  e^{\half \partial_ \eta \Bbf \partial_ \eta} e^{-\half \eta   \Abf    \eta}
  &= \frac{1}{{\rm Pf} (-\Abf)^{-1}}\, \int d\nu  \, e^{\half \nu \left(-\Abf^{-1}+ \Bbf \right)  \nu  + \  \nu   \eta} \nonumber\\ 
  &= {\rm Pf}\left(\Abf\right)\,  {\rm Pf}\left( -\Abf^{-1}+\Bbf \right) \,e^{-\half \eta\left( \frac{1}{1-\Abf \Bbf} \Abf \right) \eta}\ ,\nonumber\\ 
  &= e^{\half\Tr\ln(1-\Abf \Bbf)} e^{-\half \eta\left( \frac{1}{1-\Abf \Bbf} \Abf \right) \eta}\ ,
    \eqn{maj-iden}
\end{align}
where for the last line we used the identity ${\rm Pf} \left(\Abf \right)\,{\rm Pf} \left(\Bbf \right)= \exp\half\Tr\ln(- \Abf\Bbf)$. The above result also extends to non-invertible $\Abf$.

The Majorana result of 
\cref{eq:maj-iden} can be seen to be consistent with the Dirac result of \cref{eq:Diracident} by writing a Dirac fermion as a Majorana one with
\begin{align}
\eta = \begin{pmatrix}\chi \\ \bar\chi \end{pmatrix}\,\quad \Abf = \begin{pmatrix} 0 & -A^T \\ A & 0 \end{pmatrix},\quad \Bbf = \begin{pmatrix} 0 & B \\ -B^T & 0 \end{pmatrix}\ .
\end{align}
Then the left and right sides of \cref{eq:maj-iden} are equal to
\begin{align}
e^{\half \partial_ \eta \Bbf \partial_ \eta} e^{-\half \eta   \Abf    \eta} &= e^{  \partial_ \chi B \partial_ {\bar\chi} }e^{- \bar \chi  A   \chi}\ ,\\
e^{\half\Tr\ln(1-\Abf \Bbf)} e^{-\half \eta\left( \frac{1}{1-\Abf \Bbf} \Abf \right) \eta} &= 
e^{\Tr\log(1-AB)} \,e^{-\bar\chi\left( \frac{1}{1-AB} A \right)\chi}\  
\ ,
\end{align}
 which match the two sides of \cref{eq:Diracident}.

\bibliography{refs}

\end{document}